# IMPROVING POPULATION-SPECIFIC ALLELE FREQUENCY ESTIMATES BY ADAPTING SUPPLEMENTAL DATA: AN EMPIRICAL BAYES APPROACH

By Marc Coram and Hua Tang[1]

*Stanford University*

Estimation of the allele frequency at genetic markers is a key ingredient in biological and biomedical research, such as studies of human genetic variation or of the genetic etiology of heritable traits. As genetic data becomes increasingly available, investigators face a dilemma: when should data from other studies and population subgroups be pooled with the primary data? Pooling additional samples will generally reduce the variance of the frequency estimates; however, used inappropriately, pooled estimates can be severely biased due to population stratification. Because of this potential bias, most investigators avoid pooling, even for samples with the same ethnic background and residing on the same continent. Here, we propose an empirical Bayes approach for estimating allele frequencies of single nucleotide polymorphisms. This procedure adaptively incorporates genotypes from related samples, so that more similar samples have a greater influence on the estimates. In every example we have considered, our estimator achieves a mean squared error (MSE) that is smaller than either pooling or not, and sometimes substantially improves over both extremes. The bias introduced is small, as is shown by a simulation study that is carefully matched to a real data example. Our method is particularly useful when small groups of individuals are genotyped at a large number of markers, a situation we are likely to encounter in a genome-wide association study.

**1. Introduction.** Allele frequency at a genetic marker is one of the most important elements in studies of genetic diversity, as well as in population-based disease association studies. It plays a pivotal role in linkage studies, which model the allelic identical by descent probability, and in association studies, which directly compare the allele frequency between the affected cases and unaffected controls. Moreover, once a disease variant has

Received February 2007; revised May 2007.
[1]Supported by NIH GM073059.

been identified, accurate assessments of the allele frequency of the variant enable us to evaluate the proportion of the disease burden in a specific population that is attributable to the variant. Fueled by the recent developments in high-throughput genotyping technologies, various efforts are underway to characterize allele frequencies at a genome-wide scale in diverse populations. However, because of the still significant costs associated with these high-throughput platforms, current large-scale genomic projects often assay a large number of markers in a small number of individuals. For example, the International HapMap Project has genotyped more than four million single nucleotide polymorphisms (SNP) in 90 Africans from Nigeria (60 of which are unrelated individuals), 90 U.S. residents with northern and western European ancestry (60 of which are unrelated individuals), 45 Han Chinese from Beijing and 45 Japanese from Tokyo [International HapMap Consortium (2005)]. In another effort, Perlegen Sciences genotyped 71 Americans of European, African or Han Chinese ancestry [Hinds et al. (2005)]. The maximum likelihood estimate (MLE) of allele frequency, in this case just the observed proportion of one allele, has a binomial sampling error, which can be substantial for small samples. Small sample sizes remain a concern, even as more individuals are being genotyped, because there is a simultaneously growing concern about population stratification [Lander and Schork (1994)].

When genotypes are available from individuals representing the same populations, the allele frequency estimates can be improved by combining genotype data. On the other hand, injudicious combining of samples representing distinct populations can lead to biased estimates, as population stratification and genetic drift lead to divergence in allele frequencies among populations [Fisher (1922) and Wright (1931)]. Unfortunately, deciding whether two samples represent a homogenous population, and hence are combinable, is a delicate and subjective decision. Do the Han Chinese from Beijing (HapMap sample) and those from Los Angeles (Perlegen sample) represent the same population? Can we use the HapMap African genotypes to improve frequency estimates of Perlegen African Americans? One possible approach to address such ambiguity is a two-stage approach: one first tests whether the two samples are combinable, using a random set of markers and a procedure such as Devlin and Roeder (1999) or Pritchard and Rosenberg (1999), and, in a second stage, combine or not combine depending on the outcome of the first-stage test. This two-stage procedure, however, suffers from two potential problems. First, when only a small number of individuals have been genotyped, the first-stage test may not have sufficient power to detect the difference; on the other hand, with a sufficiently large sample size, any trivial noncongruency leads to rejection of the test, and therefore voids the possibility to combine samples. Second, the first-stage test can introduce a bias since only similar allele frequencies are allowed to be combined.

Bayesian and empirical Bayes approaches offer flexible venues for combining multiple sources of information. Lange pioneered an empirical Bayes approach for estimating allele frequencies of a single marker using data at the same marker from multiple populations [Lange (1995)]. Lockwood, Roeder and Devlin (2001) extended this approach to incorporate multi-loci genotype information via a Bayesian hierarchical model. Both methods employ a Dirichlet($\alpha$) distribution to describe the dispersion of frequencies between the different populations. The two approaches differ in how $\alpha$ is estimated: Lange's method estimates $\alpha$ by maximum likelihood at each locus separately; while Lockwood, Roeder and Devlin (2001) borrow strength across loci. These two methods are described in greater detail in Section 2.6.5. Additionally, there is a rich literature in modeling population structure and divergence using genetic polymorphism data, although the primary interests are inferences about population history and estimating parameters such as genetic distance and population size [Kitada, Hayashi and Kishino (2000), Nicholson et al. (2002) and Wilson, Weale and Balding (2003)].

In this paper we propose a new empirical Bayes approach, which offers an adaptive procedure to combine multiple samples. This method avoids the problems associated with the two-stage procedure by introducing an affinity measure, $\nu$, which is based on the global similarity between samples. There is no need to make a "hard" decision to combine or not combine, as $\nu$ parametrizes a continuous spectrum of compromise between the two extremes. As a result, our approach allows us to borrow strength from additional samples, even if they are indubitably distinct from the target population. An important advantage of our approach is that it requires no knowledge nor assumptions of the genealogy that relates various samples. As we explain in Section 2.6, our approach differs from related existing approaches in implementation as well as interpretation. We illustrate the statistical validity of our method by a series of analyses using real genotype data from HapMap and Perlegen projects; these analyses also provide some interesting biological insights regarding the populations studied by the two projects.

## 2. Method.

2.1. *Data, model and the basic idea.* Our goal is to estimate allele frequencies at a large number of bi-alleleic markers in a target population, $\mathcal{Y}$. The available data consists of $n_\mathcal{Y}$ alleles from $\mathcal{Y}$ (based on genotypes of $\frac{1}{2}n_\mathcal{Y}$ individuals), as well as a booster sample of $n_\mathcal{X}$ alleles from a (possibly related) population, $\mathcal{X}$. At each marker, we assign one allele as the "**A**" allele, and denote the observed numbers of **A** allele at marker $i$ in the two samples as $X_i$ and $Y_i$, respectively. Let $q_i$ be the frequency of **A** allele at marker $i$ in population $\mathcal{Y}$, and let $p_i$ be the corresponding frequency in $\mathcal{X}$. Within each population, we assume the Hardy–Weinberg equilibrium at each marker; furthermore, we assume that all markers are independent (i.e., in linkage equilibrium). Additionally, we assume evolutionary neutrality at the majority of markers. Some of these assumptions are not strictly necessary, as we explain in Section 2.6.3. Consequently, we model

$$X_i \sim \text{Binom}(n_{\mathcal{X}}, p_i),$$
$$Y_i \sim \text{Binom}(n_{\mathcal{Y}}, q_i) \qquad \text{(all independent)}.$$

Using genotype data from $\mathcal{Y}$ alone, the maximum likelihood estimate (MLE) of the frequency of **A** allele, $q_i$, coincides with the observed frequency,

$$\hat{q}_i = \frac{y_i}{n_{\mathcal{Y}}}.$$

Likewise, denote the MLE of the corresponding allele frequency in $\mathcal{X}$ as $\hat{p}_i$.

The empirical Bayes approach we propose is motivated by the observation that, if populations $\mathcal{X}$ and $\mathcal{Y}$ are evolutionarily related, then the allele frequencies at the corresponding markers, $p_i$ and $q_i$, are often positively associated [Jiang and Cockerham (1987)]. For example, Figure 1(a) displays a two-dimensional histogram of the MLE allele frequencies for $\sim 60{,}000$ SNPs on Chromosome 22 using the HapMap Chinese ($x$-axis) and Japanese ($y$-axis) samples. We use 15, instead of 45, Japanese individuals for this plot in order to facilitate comparison with the simulation study in Section 3.3. The higher intensity band along the diagonal indicates the high degree of association between the frequencies in JPT and CHB populations. The main statistical contribution of this paper is to develop a way to borrow strength from such association. To fix ideas, consider a marker whose JPT frequency we wish to estimate and whose CHB sample frequency happens to be 0.33. So far, what we know about this marker is that it lies somewhere in the third vertical strip in Figure 2(a). It is natural to consider the population of such markers, that is, the subset of markers whose CHB sample frequency is essentially 0.33. The third histogram shown on the right displays the corresponding allele counts in JPT. This histogram has a mean about 30 (out of 90 Japanese alleles), and the distribution is well approximated by the superimposed beta-binomial density, whose parameters were fitted by maximum likelihood. For this particular histogram, the fitted parameters are $(11.30, 22.41)$. Therefore, this data is well approximated by a model in which we consider the true frequency to have a Beta distribution and the observed counts to be Binomially distributed. Accordingly, we propose to take this Beta as an empirical prior for all of the markers in the histogram; for each marker in turn, we condition this prior on the observed JPT counts to derive a Beta posterior, and use the posterior mean as an updated frequency estimate.

2.2. *Windowed estimates.* We will formalize this basic idea by describing a windowed version of our empirical Bayes approach. The arbitrary window width $\delta$ will be removed later in a parametric version. For a single marker, a commonly used Bayesian approach assumes a Beta prior, which is a conjugate prior for the binomial distribution [Bernardo and Smith (1994)]. How-ever, a problem arises here that plagues many Bayesian approaches: how to choose the parameters, $a$ and $b$, for the prior distribution? Fortunately, with an empirical Bayes approach, we can take advantage of the large number of markers available to estimate these parameters objectively.

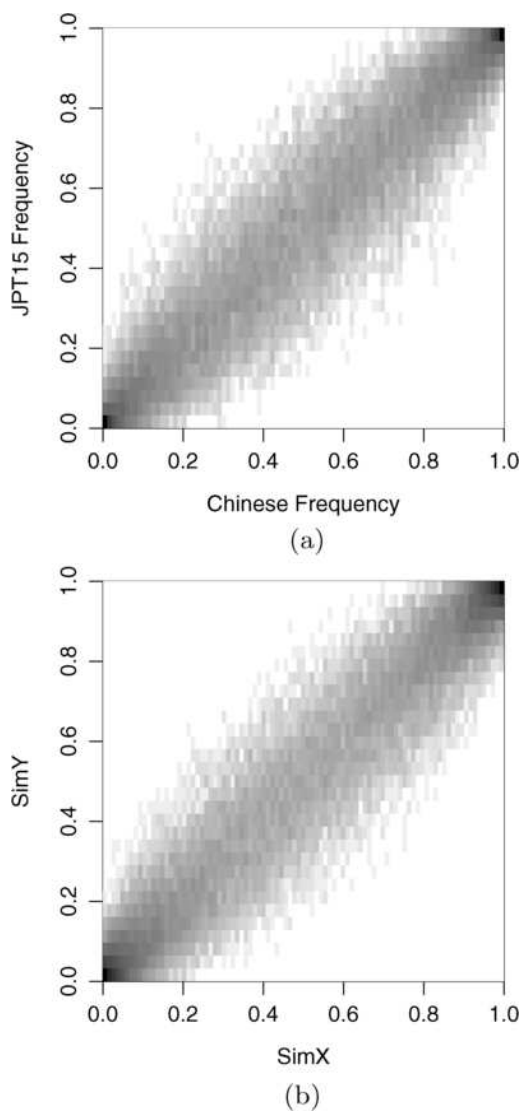

Fig. 1. *Two-dimensional Histogram of MLE of allele frequencies in* (a) 45 *HapMap Chinese individuals (x-axis) v.s.* 15 *HapMap Japanese individuals (y-axis), and* (b) 45 *simulated booster samples (x-axis) v.s.* 15 *simulated target samples (y-axis) used in Section 3.3.*

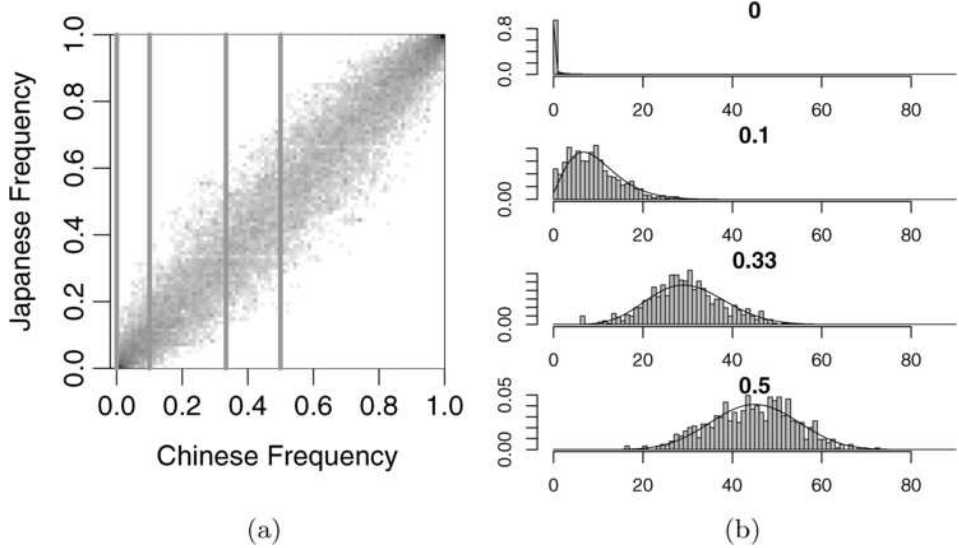

Fig. 2. *Illustration of the basic idea* (a) *scatter plot of the empirical allele frequencies in 45 HapMap Chinese individuals (x-axis) v.s. 45 HapMap Japanese individuals (y-axis);* (b) *histogram and fitted Beta-Binomial density of the allele counts in Japanese in narrow windows.*

Consider a certain marker $i$ for which we wish to estimate $q_i$. The observed frequency at marker $i$ in $\mathcal{X}$ is $\hat{p}_i$. Now consider the (presumably large) set $\mathcal{J}$ of all markers $j$ such that $\hat{p}_j$ falls within a narrow window of $\hat{p}_i$, say, $(\hat{p}_i - \delta, \hat{p}_i + \delta)$. We can understand $\mathcal{J}$ as the "population" of markers, about which we have the same information coming from population $\mathcal{X}$ as we have for the marker of interest, marker $i$. By looking at these markers as an aggregate, we can empirically determine the degree to which this $\mathcal{X}$ information does or does not inform us about marker frequency in the $\mathcal{Y}$ population. For $j \in \mathcal{J}$, we approximate the distribution $q_j$'s by a Beta distribution, Beta$(a,b)$. Under this model, the $Y_j$'s are independent and identically distributed as BetaBinom$(n_\mathcal{Y}, a, b)$. Let $\hat{a}_i$ and $\hat{b}_i$ maximize the likelihood of this data:

$$\prod_{j \in \mathcal{J}} d\text{BetaBinom}(n_\mathcal{Y}, a, b)(y_j),$$

where $d\text{BetaBinom}(n, a, b)(x)$ denotes the Beta Binomial density at $x$:

(2.1) $$\frac{\text{B}(a+x, b+n-x)}{\text{B}(a,b)} \binom{n}{x},$$

where B is the Beta function.

We therefore form the empirical Bayes prior, $q_i \sim \text{Beta}(\hat{a}_i, \hat{b}_i)$. Conditioning on $Y_i = y_i$, we obtain the posterior: $q_i | y_i \sim \text{Beta}(y_i + \hat{a}_i, n - y_i + \hat{b}_i)$. Taking the mean and variance of this posterior, we obtain the estimator

$$\hat{q}_i^{\text{EBW}} = \text{E}(q_i \mid Y_i) = \frac{y_i + \hat{a}_i}{n + \hat{a}_i + \hat{b}_i}$$

with expected squared error (ignoring the randomness in $\hat{a}_i$ and $\hat{b}_i$) of

$$\widehat{\text{Var}}(\hat{q}_i^{\text{EBW}} \mid Y_i) = \frac{\hat{q}_i^{\text{EBW}}(1 - \hat{q}_i^{\text{EBW}})}{n + \hat{a}_i + \hat{b}_i + 1}.$$

Define an affinity measure, $\nu_i = \hat{a}_i + \hat{b}_i$. In our examples, it generally happens that the mean of the empirical Bayes prior for marker $i$, $\mu_i = \frac{\hat{a}_i}{\nu_i}$ is nearly the same as $\hat{p}_i$, and has little sampling variability because of the large number of markers falling into $\mathcal{J}$. The variance of the prior can be written as $\sigma_i^2 = \mu_i(1 - \mu_i)(\nu_i + 1)^{-1}$, which decreases as $\nu_i$ increases. In other words, when $\nu_i$ is large, the booster sample exerts greater influence on the posterior estimate of $\hat{q}_i^{\text{EBW}}$. As we illustrate in the Examples, $\nu$ tends to be larger when the booster sample is biologically more closely related to the target population; hence, the procedure adaptively incorporates the booster samples.

2.3. *Parameterized estimates.* In the windowed version we selected $a$ and $b$ to maximize the likelihood of the data in each window. A more elegant approach seeks a parametric form that relates $a$ and $b$ to the conditioning information. By treating the data globally, this approach avoids the arbitrary choice of window-width. Moreover, it can readily handle the case in which we wish to simultaneously incorporate multiple booster samples (see Section 2.5). For now, continuing with the case of a single booster sample, a simple yet reasonably effective choice is EB1:

$$\begin{aligned} a(p) &= \beta_0 + \beta_1 p, \\ b(p) &= \beta_0 + \beta_1 (1-p). \end{aligned} \tag{2.2}$$

The empirical Bayes prior of $q$ with parameters, $a$ and $b$, derived from this model has a simple interpretation: "a priori" $q$ follows a Beta distribution which represents a total of $2\beta_0 + \beta_1$ pseudo-counts; a baseline of $\beta_0$ counts are assigned to each allele; the additional $\beta_1$ counts are allocated proportionally to the observed frequencies in $\mathcal{X}$.

We estimate the coefficients in (2.2), $\beta$'s, by maximizing the likelihood:

$$\prod_j d\text{BetaBinom}(n_{\mathcal{Y}}, a(\hat{p}_j), b(\hat{p}_j))(y_j),$$

where the dependency of $a$ and $b$ on the $\beta$'s is suppressed.

The model in (2.2) specifies a linear model for $a$ and $b$, and induces a symmetry condition: $a(p) = b(1-p)$. We can add higher-order terms or terms that break the symmetry, but in our analyses of real data, these terms do not contribute to significant improvements in the likelihood.

However, the likelihood does improve if the endpoints $\hat{p} = 0$ and $\hat{p} = 1$, which occur fairly often in our data, are treated as special cases. EB2 is a way to introduce additional terms that treat these cases symmetrically:

$$
\begin{aligned}
a(p) &= \beta_0 + \beta_1 p + \beta_2 \mathbf{1}_{p=0} + \beta_3 \mathbf{1}_{p=1}, \\
b(p) &= \beta_0 + \beta_1 (1-p) + \beta_2 \mathbf{1}_{p=1} + \beta_3 \mathbf{1}_{p=0},
\end{aligned}
\tag{2.3}
$$

where $\mathbf{1}$ is the indicator function.

2.4. *Spline estimates.* One might question whether the parametric forms used in the previous section impose unnecessary constraints. To allow a more flexible model of $\hat{a}$ and $\hat{b}$, we use a B-spline expansion:

$$
\begin{aligned}
a(p) &= \sum_{j=1}^{N} N_j(p) \theta_j, \\
b(p) &= \sum_{j=1}^{N} N_j(1-p) \gamma_j,
\end{aligned}
\tag{2.4}
$$

where the $N_j(x)$ are an N-dimensional set of basis functions for cubic B-splines on $[0, 1]$. We can impose the symmetry condition, $a(p) = b(1-p)$, by taking $\theta_j = \gamma_j$ for all $j$.

2.5. *Multiple boosting samples.* Several genomics projects have surveyed diverse populations. Our empirical Bayes approach generalizes naturally to such situations, incorporating empirical frequencies from several boosting samples (multiple $p$'s) in estimation of the $q$'s. For example, the parametric model EB1 for multiple boosting samples is

$$
\begin{aligned}
a(p^{(1)}, \ldots, p^{(K)}) &= \beta_0 + \sum_{k=1}^{K} \beta_k p^{(k)}, \\
b(p^{(1)}, \ldots, p^{(K)}) &= \beta_0 + \sum_{k=1}^{K} \beta_k (1 - p^{(k)}),
\end{aligned}
\tag{2.5}
$$

where $p^{(k)}$ denotes the allele frequency in boosting sample $k$. Similarly, a spline-based model is

$$
a(p^{(1)}, \ldots, p^{(K)}) = \sum_{k=1}^{K} \sum_{j=1}^{N} N_j(p^{(k)}) \gamma_j^k,
$$

(2.6)
$$b(p^{(1)}, \ldots, p^{(K)}) = \sum_{k=1}^{K} \sum_{j=1}^{N} N_j (1 - p^{(k)}) \gamma_j^k,$$

2.6. *Remarks.*

2.6.1. *An affinity measure.* For the windowed estimate, we define $\nu$ to be the median of $\hat{a} + \hat{b}$ over all windows; for EB1, we define $\nu$ as $2\hat{\beta}_0 + \hat{\beta}_1$. For the spline version, EB3, we define $\nu$ by $\int_p \hat{a}(p) + \hat{b}(p)\, dp$. In all situations, $\nu$ has a simple interpretation as the effective sample size of the boosting sample. As illustrated in the next section, $\nu$ reflects the genetic association between the two populations.

2.6.2. It can be reasoned that, if populations $\mathcal{X}$ and $\mathcal{Y}$ are essentially identical, our approach is asymptotically equivalent to pooling the genotype data directly. To verify this is how our approach behaves, we reconsider the windowed estimate.

Let $p_j \sim \text{Beta}(a_0, b_0)$ for all $j$. For marker $i$, with $\hat{p}_i = x_i/n_\mathcal{X}$, consider all markers $j$, s.t. $\hat{p}_j \in \hat{p}_i \pm \delta$. As $\delta \to 0$, this is the set of markers with $X_j = x_i$. Conditioning on $X_j = x_i$, the posterior distribution of $p_j$ is $\text{Beta}(a_0 + x_i, b_0 + n_\mathcal{X} - x_i)$. Since $\mathcal{X}$ and $\mathcal{Y}$ are identical populations, $p_j = q_j$ for all markers. It follows that $Y_j | X_j = x_i \sim \text{BetaBinom}(n_\mathcal{Y}, a_0 + x_i, b_0 + n_\mathcal{X} - x_i)$. Because the MLEs, $\hat{a}$ and $\hat{b}$, are consistent as the number of alleles becomes infinite, we have $\hat{a} \to a_0 + x_i$ and $\hat{b} \to b_0 + n_\mathcal{X} - x_i$, so that

$$\hat{q}_i^{\text{EB}} = \frac{Y_i + \hat{a}}{n_\mathcal{Y} + \hat{a} + \hat{b}}$$
$$\doteq \frac{Y_i + a_0 + x_i}{n_\mathcal{Y} + a_0 + b_0 + n_\mathcal{X}}.$$

In other words, the empirical Bayes estimator is equivalent to directly pooling data, with an additional shrinkage toward the prior pseudo proportion, $a_0/(a_0 + b_0)$.

2.6.3. In describing our approach, we have made three commonly adopted assumptions: that each marker satisfies the Hardy–Weinberg equilibrium, that pairs of markers are in linkage disequilibrium, and that the genome is under neutral evolution. The Hardy–Weinberg equilibrium (HWE) in a population requires only one generation of random mating; empirically, there is very weak evidence for systematic deviation from HWE, even in stratified (historically nonrandom mating) populations such as the Mexicans or the Puerto Ricans [Choudhry et al. (2006)]. In fact, HWE is often used as a diagnosis for genotyping error [Yonan, Palmer and Gilliam (2006)]. The assumption of linkage disequilibrium is required here so that the Beta-Binomial likelihood function can be multiplied across markers. Including tightly linked markers (i.e., correlated genotypes) in the estimation of the Beta parameters makes the effective sample size somewhat smaller than the nominal one, but should not introduce systematic bias, as long as markers are relatively evenly distributed across the genome. Finally, evolutionary neutrality warrants that the difference in allele frequencies between populations are due to genetic drift and not directional selection. Strong directional selection may create a situation in which allele frequencies are similar in two populations at most loci, except at a few loci where selection results in large allele frequency discrepancy. While there is evidence that various parts of the human genome have been subjected to recent positive selection [Sabeti et al. (2006) and Voight et al. (2006)], the selection coefficients are likely low [Kimura (1968)]. Further, evolutionary neutrality holds in a large proportion of the genome, in particular, noncoding SNPs and synonymous SNPs. Besides, in the unlikely scenario that strong selection leads to divergence between two populations at a genome-wide scale, our approach will not produce severely biased frequency estimates, because the affinity measure will be low and, therefore, the booster data are not allowed to influence the estimates substantially.

2.6.4. One may question the appropriateness of the Beta prior we assume, which is conveniently the conjugate prior for a Binomial distribution [Skellam (1948)]. Under selective neutrality and for populations that have reached an mutation-drift equilibrium, Wright (1951) showed that the allele frequencies at bi-allelic loci follow a beta distribution. Visual inspection indicates that the empirical allele frequencies of the HapMap European samples follow a Beta distribution reasonably well, although there are slight excess rare alleles, that is, frequencies near 0 or 1. While it may be possible for a different prior to somewhat improve the fit, it is difficult to find a perfect prior. This is because the SNPs genotyped in a project seldom represent a random sample of all polymorphic sites, and the ascertainment bias distorts the underlying frequency spectrum. When the ascertainment procedure is known, it is possible to correct the bias [Nicholson et al. (2002) and Nielsen, Hubisz and Clark (2004)]. Unfortunately, the ascertainment schemes are often so complex that it is difficult to correct for the bias [Clark et al. (2005)].

2.6.5. We are now in a good position to explain the difference between our methods and the empirical Bayes approach of Lange (1995), which aims to improve the frequency estimate at a single marker using genotypes from multiple populations. This method models the allele frequencies at the marker in various population as independent draws from a single prior distribution, which is chosen to maximize the likelihood of the observed allele counts in all populations. The posterior mean represents a shrinkage toward the pooled population average. In contrast, our method is closer to a regression model; the frequencies in the target population are modeled conditionally on the boosting population frequencies. Thus, it borrows strength from the information in the frequencies in the boosting population to better estimate their corresponding frequencies in the target. Yet only those populations whose frequencies are thought to be informative about the frequencies in the target are used. It considers the set of all markers with a given frequency in the boosting population to be the collection of markers that captures the pertinent conditioning information.

The hierarchical Bayesian approach taken by Lockwood, Roeder and Devlin (2001) can be expected to behave like our estimator in many respects, although coming to this estimate from a different direction. Specifically, the hierarchical structure models the population-specific allele frequency by a locus-specific Dirichlet distribution, so that the sum of the Dirichlet parameters controls the divergence between populations, and is theoretically motivated by Wright's $F_{st}$ [Wright (1951)]. The Dirichlet parameters are allowed to vary across loci, but the hierarchical structure of the model borrows strength from all loci to determine, roughly speaking, the overall degree of divergence. At each locus, the allele frequency in each population is modeled as a "symmetric" departure from that of an implicit ancestor. The specific form is reasonable so long as the populations are related through a star-shaped phylogeny. However, it would appear inappropriate to treat the populations in this symmetric manner if some of the populations are more closely related than others. In contrast, our approach does not impose an underlying population history model. Instead, it endeavors to be flexible and data-adaptive for the purpose of estimating allele frequencies in the target population and estimating the effective genetic association between the target and each of the booster samples.

2.6.6. In the examples below, we "orient" our data so that the "major" allele whose frequency we are estimating corresponds to the alphabetically lesser nucleotide (A<C<G<T) as it would occur on the positive strand of the chromosome. As this orientation treats all markers equitably, it is not too surprising that the frequencies, thus defined, are quite symmetrically distributed about $\frac{1}{2}$. If the data is oriented based on other information, for example, on the basis of the allele present in the reference sequence of the human genome, the symmetry is broken. It is also more "neutral." For example, if an oracle were to orient the data so that the allele that is more likely in, say, Europeans was always the major allele, this actually provides extra information (e.g., you would know not to estimate a frequency below 0.5 for Europeans on any allele), but this does not treat different population groups equitably (e.g., African population frequencies wouldn't obey this rule). The fair way to introduce this extra information, we argue, is to actually provide the genotype data that your orientation is based on as a (potential) booster sample.

**3. Results.** In this section we examine the performance of the empirical Bayes estimates using genotype data collected from the International HapMap Project [International HapMap Consortium (2005)] and by the Perlegen Sciences [Hinds et al. (2005)]. For the HapMap data, we used the genotype data (Public Release #21) from unrelated individuals representing four ethnic populations; these include the following: 60 Yoruba in Ibadan, Nigeria (YRI), 60 U.S. residents with ancestry from northern and western Europe (CEU), 45 Han Chinese in Beijing, China (CHB), and 45 Japanese in Tokyo, Japan (JPT). Data from the Perlegen Project includes 24 European Americans, 23 African Americans and 24 Han Chinese from the Los Angeles area. Because our goal is to develop this statistical approach, we only used SNPs on chromosome 22, leaving $\sim 55{,}000$ SNPs in the HapMap data and $\sim 20{,}000$ SNPs in the Perlegen data, of which $\sim 14{,}800$ overlap.

3.1. *Booster sample and target sample represent identical populations.* Our first experiment examines the performance of the empirical Bayes estimates, when the populations represented by the target sample and the booster sample coincide. To do so, we randomly split each population sample in HapMap into two sets of approximately equal size, and treat one set as the target sample and the other as a booster. Intuitively, if we knew that the two samples came from the same population, the most efficient frequentist estimator is to simply compute the MLE using the pooled genotype data. The equivalent empirical Bayes estimator would use $\hat{a}_i = x_i$ and $\hat{b}_i = n_\mathcal{X} - x_i$. In other words, the prior would use the observed allele counts in the booster sample as the pseudo-counts. Table 1 summarizes the results for each population, demonstrating that the empirical Bayes approaches, both EB1 and EB2, approximately recover the pooling estimator on this data. For example, in terms of EB1, the "pooling-equivalent" empirical Bayes prior uses $\beta_0 = 0$ and $\beta_1 = n_\mathcal{X}$ so that $\hat{a}(\hat{p}_i) = 0 + n_\mathcal{X} \hat{p}_i$ and $\hat{b}(\hat{p}_i) = 0 + n_\mathcal{X}(1 - \hat{p}_i)$. As shown in Table 1, the estimated coefficients ($\hat{\beta}_1$) are very close to $n_\mathcal{X}$ so that EB1 and EB2 do indeed behave much like pooling on this example, as one would hope. Instead of using 0 baseline pseudo-counts, though, the likelihood criterion selects small $\beta_0$ values near 0.10. This is sensible since a $\text{Beta}(0.1, 0.1)$ distribution approximates the unconditional distribution of allele frequencies in these populations.

EMPIRICAL BAYES ESTIMATE OF ALLELE FREQUENCY 13

TABLE 1
*Parametric estimators for HapMap populations, when the target and booster populations coincide. EB1 refers to the model in (2.2); EB2 refers to the model in (2.3)*

| pop | $n_\mathcal{Y}$ | $n_\mathcal{X}$ | EB1 | | EB2 | | | |
|---|---|---|---|---|---|---|---|---|
| | | | $\beta_0$ | $\beta_1$ | $\beta_0$ | $\beta_1$ | $\beta_2$ | $\beta_3$ |
| YRI | 60 | 60 | 0.13 | 58.19 | 0.63 | 58.62 | $-0.55$ | $-26.26$ |
| CEU | 60 | 60 | 0.11 | 58.90 | 0.59 | 58.87 | $-0.51$ | $-19.64$ |
| CHB | 44 | 46 | 0.10 | 46.36 | 0.53 | 46.27 | $-0.46$ | $-13.62$ |
| JPT | 44 | 46 | 0.08 | 44.60 | 0.65 | 44.69 | $-0.60$ | $-17.97$ |

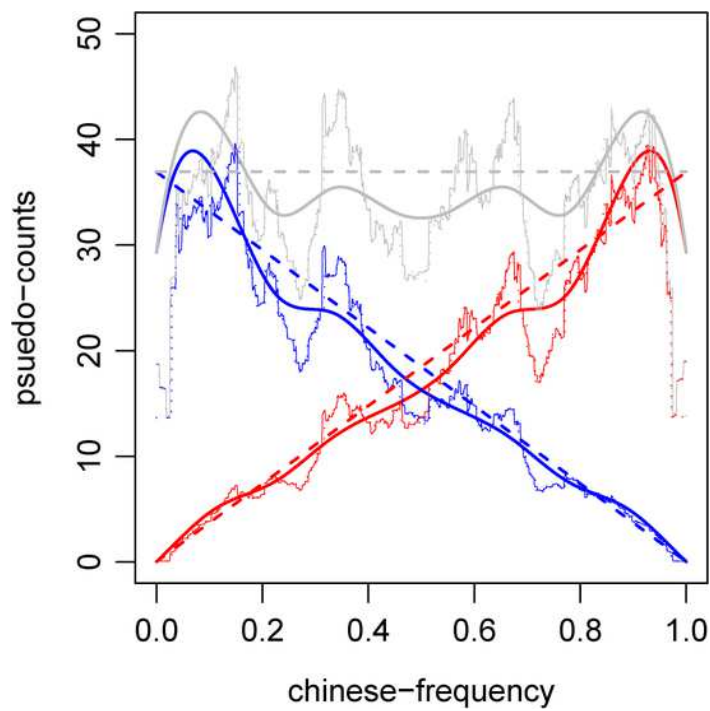

FIG. 3. *Estimated parameters for the empirical Bayes priors as a function of the observed frequency in the booster sample. Target sample consists of 15 HapMap Japanese individuals; booster sample consists of 45 HapMap Chinese individuals. Blue lines are parameters a in the Beta prior; red lines are b; gray lines are a + b, which is a local measure of affinity between the target and booster populations. Heavy solid lines are parameters for spline model, EB3; thin lines are parameters for windowed implementation; dotted lines are parameters for linear model EB1.*

3.2. *HapMap Chinese boosts HapMap Japanese.* In the second experiment we use the HapMap Chinese sample (CHB) to boost frequency esti-mates for the Japanese (JPT). As a "validation" sample, we set aside 30 JPT individuals, and apply empirical Bayes estimators on 15 JPT and 45 CHB. Figure 3 plots the estimated parameters $\hat{a}$ and $\hat{b}$ using windowed, parametric and spline variations of the empirical Bayes estimators. Visual inspection suggests that these different models produce similar empirical Bayes priors. For the linear estimator, EB1, the estimated affinity measure is $\nu = 36.96$ ($\beta_0 = 0.038, \beta_1 = 36.88$) substantially lower than the nominal 90 alleles. Since the previous example suggests that the empirical Bayes estimators will approximate direct pooling when the boosting sample represents the same population as the target, the discrepancy between empirical Bayes estimators and direct pooling indicates a nontrivial genetic heterogeneity between the Chinese and Japanese. Pooling these samples would not be justified, but the high affinity means that our method can still borrow a large amount of strength.

Since our goal is to improve the Japanese allele frequency estimates, we compute the mean squared error (MSE) for the various estimators, treating the MLE of allele frequencies in the "validation" sample as an *imperfect* gold standard. As we explain in the Appendix, calculating the MSE by treating the validation sample as a perfect gold standard produces an upward bias, which can be corrected. In what follows we report the corrected MSE. Using the MLE of allele frequencies on the 15 JPT alone, the MSE is $2.83 \times 10^{-3}$. In contrast, EB3 produces an MSE of $1.273 \times 10^{-3}$, achieving a 55% error reduction without any a priori assumption of population homogeneity. The EB1 model fits essentially just as well (MSE of $1.279 \times 10^{-3}$), itself only slightly better than the windowed estimate (MSE of $1.282 \times 10^{-3}$). Accordingly, we favor the simple and relatively interpretable EB1 model for general application.

In this case, if one simply treats the Chinese and Japanese individuals as sampled from a homogeneous population and computes the MLE on the pooled sample of 60 individuals, the corrected MSE is reduced to $1.51 \times 10^{-3}$, not so much higher than EB3. However, it does not always work out this well for the pooled estimate: if the populations are less closely associated, the bias introduced from pooling can overwhelm the variance reduction (e.g., if Europeans are pooled in with African Americans to "improve" their frequency estimates; see Section 3.3.1). Moreover, in practice, few investigators would feel comfortable pooling heterogeneous populations such as Chinese and Japanese. Our method has the advantage that it automatically obtains the right degree of pooling, and thus allows us to borrow information from a booster sample *even when we know the samples represent heterogeneous populations.*

3.3. *Simulated data.* In order to accurately assess the MSE, bias and variance in a situation where we have a true gold standard, we perform a simulation experiment. The data consists of 55,000 markers, whose frequencies in $\mathcal{X}$, $p$, are drawn independently from Beta(0.198, 0.198); conditional on $p$, the corresponding frequencies in $\mathcal{Y}$, $q$, are drawn from Beta($a(p_i), b(p_i)$), where $a$ and $b$ follows the model of (2.2). The coefficients, $\beta$'s, used in (2.2) are chosen so that the joint distribution of $\hat{p}$ and $\hat{q}$ approximates that of the HapMap Chinese and Japanese. Given $p_i$ and $q_i$, we next generate genotype data by sampling $Y_i \sim \text{Binom}(30, q_i)$, and $X_i \sim \text{Binom}(90, p_i)$. The scatter plot of the simulated data is shown in Figure 1(b), and resembles the Chinese v.s. Japanese plot [Figure 1(a)] in both marginal and joint distributions.

Figure 4 displays the estimated $a$ and $b$ using various methods. We see that they are qualitatively similar. To compute MSE, bias and variance, we use the underlying frequencies in population $\mathcal{Y}$ as the gold standard. The MSE using the 30 observed alleles from the target sample is $2.3 \times 10^{-3}$; by directly pooling all 120 observed alleles, the MSE is $1.2 \times 10^{-3}$, while the MSE using EB3 is $1.0 \times 10^{-3}$. Figure 5 shows that the MSE using the boosting sample (red curve) can be substantially lower than the MLE using samples from $\mathcal{Y}$ alone (black curve), especially for frequencies near 0.5. The bias-variance decomposition, also shown, indicates that the bias introduced by the empirical Bayes procedure is quite small.

3.3.1. *Admixed populations.* Estimating allele frequencies in an admixed population offers an interesting application of our methods. An admixed population arises when reproductively isolated ancestral populations mate, producing offsprings whose genome represents a mixture of alleles from multiple ancestral populations. Two of the largest minority populations in the U.S. are both recently admixed: African Americans are largely an African group with recent European admixture, and the Hispanics represent various degrees of mixing among Native Americans, Europeans and Africans. Increasing numbers of genetics studies are focusing on one of these admixed populations. Intuitively, the frequencies in the admixed population (e.g., African Americans) resemble a weighted average of the corresponding frequencies in the ancestral populations (e.g., Europeans and Africans). Therefore, existing genotype data on the ancestral populations should provide information on the allele frequencies in an admixed populations.

Here we illustrate this idea using the HapMap YRI and CEU samples to improve African American allele frequencies estimated from the Perlegen data. We use 12 of the Perlegen African American individuals to estimate $\hat{q}$, while reserving the other 12 individuals as an *imperfect* gold standard. The standard and common practice, which uses the 12 African Americans alone, produces an MSE of $13.2 \times 10^{-3}$. There are several ways to incorporate HapMap samples: using CEU alone, EB1 estimates $(\beta_0, \beta_1)$ to be $(0.66, 4.45)$; using YRI alone, EB1 estimates $(\beta_0, \beta_1)$ to be $(1.48, 34.29)$. These results indicate that the YRI population has higher affinity to the

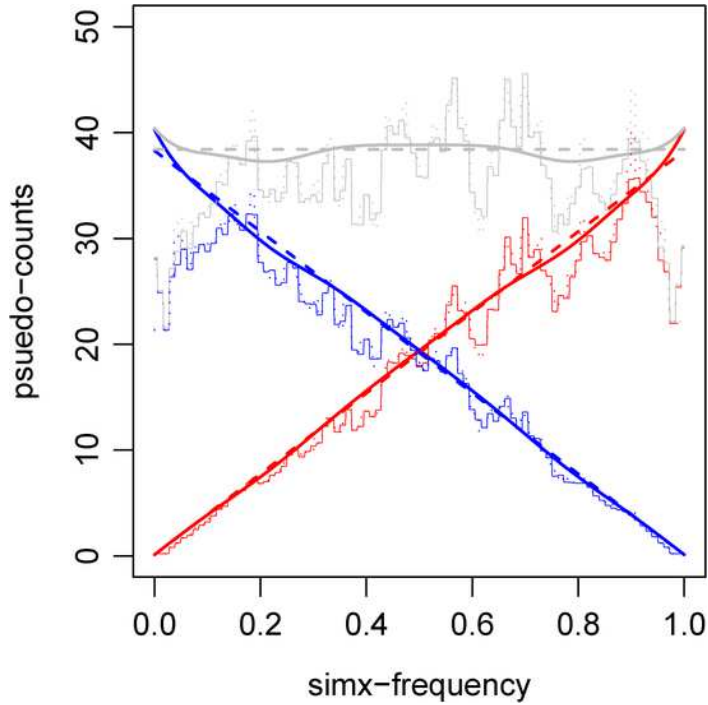

Fig. 4. *Estimated parameters for the empirical Bayes priors as a function of the observed frequency in the booster sample. The target and booster samples consist of* 30 *and* 90 *simulated alleles, respectively. Blue lines are parameters a in the Beta prior; red lines are b; gray lines are a + b, which is a local measure of affinity between the target and booster populations. Heavy solid lines are parameters for spline model, EB3; thin lines are parameters for windowed implementation; dotted lines are parameters for linear model EB1.*

African Americans. It is important to note that naively pooling the CEU and the African American samples will *increase* the MSE. When we have booster samples from both YRI and CEU, our method of choice is model (2.5), which allows us to incorporate YRI and CEU simultaneously, the parameter estimates are ($\beta_0 = 1.22, \beta_{\text{YRI}} = 64.37, \beta_{\text{CEU}} = 18.90$). Interestingly, the fraction, $\frac{\beta_{\text{CEU}}}{\beta_{\text{YRI}} + \beta_{\text{CEU}}} = .23$, resembles previously estimated European ancestry in the African Americans in the literature [Parra et al. (1998)]. The parameter estimates and the MSE evaluated using the remaining 12 African Americans are shown in Table 2.

This example highlights the advantage of the empirical Bayes method over MLE based on pooling: first, if we had only a European booster sample, we would be better off not using it at all than naively pooling it. The empirical Bayes method estimates a very low affinity of 5.65; between the two extremes, this is closest to ignoring the booster sample. Yet it does not

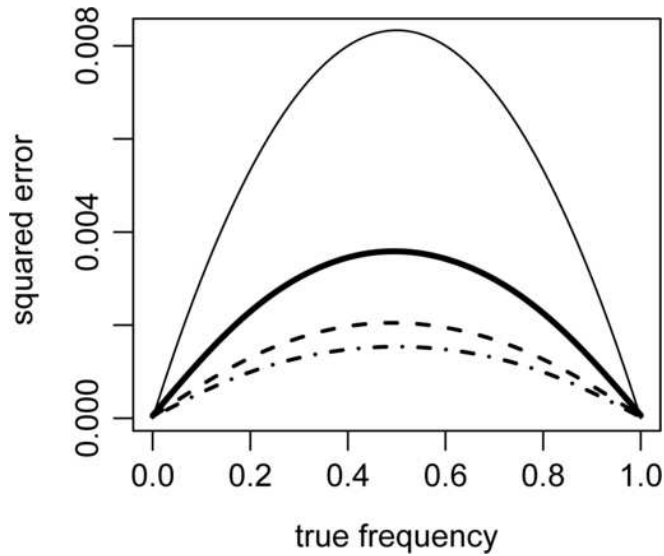

Fig. 5. *MSE (thick solid), bias$^2$ (dashed) and variance (dash-dot) of the empirical Bayes estimates on the simulated data presented in Section* 3.3. *The thin line represents the expected MSE of the MLE using the target sample alone.*

ignore the CEU sample completely, thereby still achieving an 18% reduction in MSE. Second, no matter what the booster sample is (CEU alone, YRI alone, or CEU and YRI), EB achieves a smaller MSE than either naively pooling or ignoring the same booster sample(s). Finally, we note that the influence of both CEU and YRI are greater in the presence of two booster samples than when there is only one booster sample, and the minimum MSE is achieved using two booster samples simultaneously. Loosely speaking, we

Table 2

*Coefficient estimates and MSE comparison for estimating frequencies in African Americans. Target sample (column Y) consists of* 12 *Perlegen African American individuals; booster samples (column X) consist of* 60 *CEU and* 60 *YRI individuals from HapMap. MSEs are computed using* 12 *independent African American individuals as a gold standard, with the bias correction described in the Appendix. The column labeled* $\text{MSE}_{\text{MLE}}$ *is based on naively pooling genotype data from samples X and Y, regardless of population origin. The column labeled* $\text{MSE}_{\text{EB}}$ *is based on equation (2.5)*

| X | Y | $\beta_0$ | $\beta_{\text{CEU}}$ | $\beta_{\text{YRI}}$ | $\text{MSE}_{\text{MLE}}$ | $\text{MSE}_{\text{EB}}$ |
|---|---|---|---|---|---|---|
|  | Af. Am | — | — | — | $6.24 \times 10^{-3}$ |  |
| CEU | Af. Am | 0.65 | 4.34 | — | $22.48 \times 10^{-3}$ | $5.13 \times 10^{-3}$ |
| YRI | Af. Am | 1.47 | — | 34.51 | $3.35 \times 10^{-3}$ | $2.47 \times 10^{-3}$ |
| YRI and CEU | Af. Am | 1.22 | 18.90 | 64.37 | $4.15 \times 10^{-3}$ | $1.15 \times 10^{-3}$ |

strike the right balance in weighing the booster samples, and at the same time extract more information from each relevant booster sample. Most importantly, we achieve such balance and optimality without assuming any known genetic relationship among the target and the booster samples.

**4. Discussion.** Estimating allele frequencies is a basic yet important step in many genetic studies. For example, population-specific allele frequency is the essential source of information used in a series of analyses, which inferred genetic structure, as well as correlation between spatial pattern of genetic variation and geography, in the Human Genome Diversity Project—Centre d'Etude du Polymorphisme Humain (HGDP-CEPH) Human Genome Diversity Panel [Ramachandran et al. (2005) and Rosenberg et al. (2002, 2005)]. The HGDP-CEPH consists of 1000 individuals representing 52 world-wide populations, many of which are represented by 10–25 individuals. In such small samples, population-specific allele frequency estimates are subject to substantial sampling errors; as a result, the inferred genetic clustering pattern was unstable especially for populations with small samples [Rosenberg et al. (2005)]. In linkage analysis of extended pedigrees, where not all relevant members of a pedigree have been genotyped, population allele frequency is required to infer identity by descent (IBD) information [Risch (1900), Weeks and Lange (1998) and Lockwood, Roeder and Devlin (2001)].

Likewise, accurate allele frequency estimates play an important role in a whole-genome case-control association study (WGA), which compares the allele frequency between a group of affected cases and a group of unrelated healthy controls. Numerous WGA are underway, with a sample size on the order of thousands of subjects, each genotyped at 100,000–500,000 SNPs. Because of the need to correct for multiple comparisons, large samples are required for detecting risk alleles that are rare or have moderate effect [Hirschhorn and Daly (2005) and Wang et al. (2005)]. While in most ongoing case-control studies, the overall sample size is chosen to achieve good statistical power, stratified analyses are often performed on much smaller subsets of participants representing minority groups. Thus, estimating allele frequency from a small sample size remains a concern in practice.

In this paper we proposed an empirical Bayes approach, which enabled us to improve population-specific allele frequency estimates by adaptively incorporating genotype data from related populations. The flexibility and computational efficiency of our approach allows it to be incorporated in existing genetic data analyses. In the context of case-control association studies, although the approach we propose not directly address the hypothesis testing problem, it can be further developed to do so. Powered by the new generation of high-throughput platforms, we expect a bloom in genotype data from diverse populations. When genotypes from additional unaffected individuals are available from an external source, we can apply our method and reduce the uncertainty in the frequencies of the control samples.

Our approach differs from existing approaches in several important aspects. First, since our goal is to improve the allele frequency estimates in a specific target population, we do not treat all populations symmetrically. Of course, if we want to improve the estimates in all populations, we can simply apply our method to target each population in turn. Second, by concentrating on allele frequency estimation, our method does not require assuming an underlying genealogy or a common ancestral population. This is attractive because genealogy (or coalescent)-based approaches either are restricted to analysis of data from a genetic region with negligible recombination (such as Y-chromosome or mtDNA), or require heavy computation on elaborate ancestral recombination graphs [Nordborg (2001)]. Thus, coalescent-based approaches are not easily applicable to data of genomic scale. Likewise, modeling a common ancestral population either requires a full genealogical approach as described above, or making simplifying assumptions such as a star-shaped genealogy. In contrast, by avoiding explicit modeling of full population history, our approach is not only computationally efficient, but also is more robust to unknown and complex demographic history.

Throughout this paper we have considered estimation with respect to a squared- error loss. In practice, one might be more concerned, say, about proportional errors in frequency; taken to the extreme, this requires that special attention be paid to rare alleles. In the examples we consider, we find that a single beta-binomial model fits the sampling distribution of the target frequencies reasonably well. So long as this remains the case, one can continue to compute the posterior in the manner we describe but may wish to consider, for example, the posterior median as an alternative to the posterior mean. By this choice, one minimizes the posterior L1 loss on both the frequency as well as the log-odds scales. On the other hand, if it turns out that the sampling distribution of rare allele's frequencies are not adequately approximated by a beta-binomial, it would be sensible to consider a generalization of this work in which, instead of fitting family of parameterized beta distributions, one attempted to fit a mixture of beta's, so that the extra component can make a more refined model for the appearance of rare alleles.

Our examples using real genetic data suggest that incorporating additional boosting samples can often substantially improve the frequency estimates by introducing only a small degree of bias in exchange for the variance reduction. However, the improvement in the estimation describes the behavior of the estimate averaged over all SNPs. There may be a small number of SNPs whose allele frequencies differ substantially between populations. Therefore, as a word of caution, we point out that the approach we propose here may not be appropriate for some applications. For example, if one's goal is to detect SNPs whose frequencies differ significantly between two populations, then using each population to boost the other tends to shrink the overall allele frequency difference. In future research, we plan to develop extensions to detect such SNPs.

Empirical Bayes approaches, pioneered by Robbins (1964), offer a natural and unified framework for incorporating auxiliary information. We hope that the promising results we report here will inspire further development for analyzing the impending large genotyping studies.

## APPENDIX

In our real-data examples the MSE is computed using a gold standard that is imperfect; we now show that this results in an upward bias compared to the true MSE. We can estimate the bias and then subtract it to yield unbiased MSE estimates. Let $\hat{q}$ be the estimator whose MSE we are estimating, let $\hat{q}^{\text{val}}$ denote the allele frequency in an *independent* validation sample (imperfect gold standard), and let $\widetilde{MSE} = \frac{1}{N}\sum_i (\hat{q}_i - \hat{q}_i^{\text{val}})^2$ denote the associated MSE estimate using the imperfect gold standard. Under the assumed statistical model, the $\hat{q}_i^{\text{val}}$'s are independent of the data used for estimation, and are unbiased estimates of the true frequencies $q_i$. To compute the bias of $\widetilde{MSE}$, add and subtract $q_i$, expand the square, and take expectations:

$$\mathbf{E}\widetilde{MSE} = \mathbf{E}\frac{1}{N}\sum_i (\hat{q}_i - q_i + q_i - \hat{q}_i^{\text{val}})^2,$$

$$(A.1) \qquad = \mathbf{E}\frac{1}{N}\sum_i [(\hat{q}_i - q_i)^2 + (\hat{q}_i^{\text{val}} - q_i)^2 + 2(\hat{q}_i - q_i)(q_i - \hat{q}_i^{\text{val}})],$$

$$= \mathbf{E} MSE + \frac{1}{N}\sum_i \mathbf{Var}(\hat{q}_i^{\text{val}}),$$

where the cross term in (A.1) has expectation zero, being a product of independent factors, of which the second has expectation 0 because $\hat{q}_i^{\text{val}}$ is unbiased. The variance of $\hat{q}_i^{\text{val}}$ is $q_i(1-q_i)/n_{\text{val}}$, due to binomial variation. Attempting a plug-in estimate for this variance, we observe that $\mathbf{E}\hat{q}_i^{\text{val}}(1-\hat{q}_i^{\text{val}}) = q_i - \mathbf{E}(\hat{q}_i^{\text{val}})^2 = q_i - q_i^2 - [\mathbf{E}(\hat{q}_i^{\text{val}})^2 - q_i^2] = q_i(1-q_i) - \mathbf{Var}(\hat{q}_i^{\text{val}}) = q_i(1-q_i)(1 - \frac{1}{n_{\text{val}}})$. Therefore, an unbiased estimate of $\mathbf{Var}(\hat{q}_i^{\text{val}})$ is $\hat{q}_i^{\text{val}}(1-\hat{q}_i^{\text{val}})/(n_{\text{val}} - 1)$. Consequently, an unbiased estimate of the MSE is

$$\widetilde{MSE} - \frac{1}{N}\sum_i \frac{\hat{q}_i^{\text{val}}(1-\hat{q}_i^{\text{val}})}{n_{\text{val}} - 1}.$$

| | |
|---|---|
| DEPARTMENT OF HEALTH RESEARCH AND POLICY | DEPARTMENT OF GENETICS |
| STANFORD UNIVERSITY | STANFORD UNIVERSITY SCHOOL OF MEDICINE |
| STANFORD, CALIFORNIA 94305 | STANFORD, CALIFORNIA 94305 |
| USA | USA |
| E-MAIL: mcoram@stanford.edu | E-MAIL: huatang@stanford.edu |